\documentclass{article}
\usepackage{LaThuileFPSpro}
\catcode`@=12
\def\be{\begin{equation}}
\def\ee{\end{equation}}
\def\bea{\begin{eqnarray}}
\def\eea{\end{eqnarray}}
\def\beq{\begin{equation}}
\def\eeq{\end{equation}}
\def\bq{\begin{quote}}
\def\eq{\end{quote}}
\def\gappeq{\mathrel{\rlap {\raise.5ex\hbox{$>$}} {\lower.5ex\hbox{$\sim$}}}}
\def\lappeq{\mathrel{\rlap{\raise.5ex\hbox{$<$}} {\lower.5ex\hbox{$\sim$}}}}

\def\epm#1#2{\hbox{${\lower1pt\hbox{$\scriptstyle +#1$}}
\atop {\raise1pt\hbox{$\scriptstyle -#2$}}$}}

\def\gsim{\mathrel{\rlap{\lower4pt\hbox{\hskip1pt$\sim$}}
    \raise1pt\hbox{$>$}}}         

\def\frac#1#2{{{#1}\over {#2}}}

\def\GeV{{\rm GeV}}

\def\bq{\bar{q}}
\catcode`@=11 
\def\slash#1{\mathord{\mathpalette\c@ncel#1}}
 \def\c@ncel#1#2{\ooalign{$\hfil#1\mkern1mu/\hfil$\crcr$#1#2$}}
\def\lsim{\mathrel{\mathpalette\@versim<}}
\def\gsim{\mathrel{\mathpalette\@versim>}}
 \def\@versim#1#2{\lower0.2ex\vbox{\baselineskip\z@skip\lineskip\z@skip
       \lineskiplimit\z@\ialign{$\m@th#1\hfil##$\crcr#2\crcr\sim\crcr}}}
\catcode`@=12 

\def\beq{\begin{equation}}
\def\eeq{\end{equation}}
\def\bea{\begin{eqnarray}}
\def\eea{\end{eqnarray}}
\def\bet{\begin{tabular}}
\def\eet{\end{tabular}}
\def\bes{\begin{subequations}\bea}
\def\ees{\eea\end{subequations}}

\def\e{\epsilon}

\def\be{\begin{equation}}
\def\ee{\end{equation}}
\def\bc{\begin{center}}
\def\ec{\end{center}}
\def\bea{\begin{eqnarray}}
\def\eea{\end{eqnarray}}
\def\dd{\displaystyle}
\def\nn{\nonumber}

\catcode`@=11
\def\marginnote#1{}
\newcount\hour
\newcount\minute
\newtoks\amorpm
\hour=\time\divide\hour by60
\minute=\time{\multiply\hour by60 \global\advance\minute by-\hour}
\edef\standardtime{{\ifnum\hour<12 \global\amorpm={am}%
        \else\global\amorpm={pm}\advance\hour by-12 \fi
        \ifnum\hour=0 \hour=12 \fi
        \number\hour:\ifnum\minute<10 0\fi\number\minute\the\amorpm}}
\edef\militarytime{\number\hour:\ifnum\minute<10 0\fi\number\minute}
\def\draftlabel#1{{\@bsphack\if@filesw {\let\thepage\relax
   \xdef\@gtempa{\write\@auxout{\string
      \newlabel{#1}{{\@currentlabel}{\thepage}}}}}\@gtempa
   \if@nobreak \ifvmode\nobreak\fi\fi\fi\@esphack}
        \gdef\@eqnlabel{#1}}
\def\@eqnlabel{}
\def\@vacuum{}
\def\draftmarginnote#1{\marginpar{\raggedright\scriptsize\tt#1}}
\def\draft{\oddsidemargin 0.0truein
        \def\@oddfoot{\sl preliminary draft \hfil
        \rm\thepage\hfil\sl\today\quad\militarytime}
        \let\@evenfoot\@oddfoot \overfullrule 3pt
        \let\label=\draftlabel
        \let\marginnote=\draftmarginnote
   \def\@eqnnum{(\theequation)\rlap{\kern\marginparsep\tt\@eqnlabel}%
\global\let\@eqnlabel\@vacuum}  }
\catcode`@=12
\def\be{\begin{equation}}
\def\ee{\end{equation}}
\def\bea{\begin{eqnarray}}
\def\eea{\end{eqnarray}}
\def\beq{\begin{equation}}
\def\eeq{\end{equation}}
\def\bq{\begin{quote}}
\def\eq{\end{quote}}
\def\gappeq{\mathrel{\rlap {\raise.5ex\hbox{$>$}} {\lower.5ex\hbox{$\sim$}}}}
\def\lappeq{\mathrel{\rlap{\raise.5ex\hbox{$<$}} {\lower.5ex\hbox{$\sim$}}}}

\def\epm#1#2{\hbox{${\lower1pt\hbox{$\scriptstyle +#1$}}
\atop {\raise1pt\hbox{$\scriptstyle -#2$}}$}}

\def\gsim{\mathrel{\rlap{\lower4pt\hbox{\hskip1pt$\sim$}}
    \raise1pt\hbox{$>$}}}         

\def\frac#1#2{{{#1}\over {#2}}}

\def\GeV{{\rm GeV}}

\def\bq{\bar{q}}
\catcode`@=11 
\def\slash#1{\mathord{\mathpalette\c@ncel#1}}
 \def\c@ncel#1#2{\ooalign{$\hfil#1\mkern1mu/\hfil$\crcr$#1#2$}}
\def\lsim{\mathrel{\mathpalette\@versim<}}
\def\gsim{\mathrel{\mathpalette\@versim>}}
 \def\@versim#1#2{\lower0.2ex\vbox{\baselineskip\z@skip\lineskip\z@skip
       \lineskiplimit\z@\ialign{$\m@th#1\hfil##$\crcr#2\crcr\sim\crcr}}}
\catcode`@=12 

\begin{document}

\title{ NORMAL AND SPECIAL MODELS OF NEUTRINO MASSES AND MIXINGS
}
\author{
  Guido Altarelli        \\
  {\em CERN, Department of Physics, Theory Division}\\ 
{CH-1211 Geneva 23, Switzerland} \\
and\\
  {\em Dipartimento di Fisica `E.~Amaldi', Universit\`a di Roma Tre}\\
{INFN, Sezione di Roma Tre, I-00146 Rome, Italy}
  }
  
\maketitle

\baselineskip=11.6pt

\begin{abstract}
  One can make a  distinction between "normal" and "special" models. For normal models  $\theta_{23}$ is not too close to maximal and $\theta_{13}$ is not too small, typically a small power of the self-suggesting order parameter $\sqrt{r}$, with $r=\Delta m_{sol}^2/\Delta m_{atm}^2 \sim 1/35$. Special models are those where some symmetry or dynamical feature assures in a natural way the near vanishing of $\theta_{13}$ and/or of $\theta_{23}- \pi/4$. Normal models are conceptually more economical and much simpler to construct. Here we focus on special models, in particular a recent one based on A4 discrete symmetry and extra dimensions that leads in a natural way to a Harrison-Perkins-Scott mixing matrix.
\end{abstract}
\vspace{1.5cm}
\begin{center}
To be published in the Proceedings of\\
Rencontres de Physique de la Vall\' ee D'Aoste, La Thuile, Italy\\
27 February -- 5 March 2005
\end{center}
\vspace {4.0cm}
\begin{flushleft}
CERN-PH-TH/2005-140\\
27 July 2005
\end{flushleft}
\newpage

\section{Introduction}

By now there is convincing evidence for solar and atmospheric neutrino oscillations. The $\Delta m^2$ values and mixing angles are known with fair accuracy.  A summary, taken from Ref. \cite{one} of the results is shown in Table~\ref{tab01}. For the $\Delta m^2$ we have:  $\Delta m^2_{atm}\sim 2.5~10^{-3}~eV^2$ and  $\Delta m^2_{sol}\sim ~10^{-5}~eV^2$. As for the mixing angles, two are large and one is small. The atmospheric angle $\theta_{23}$ is large, actually compatible with maximal but not necessarily so: at $3\sigma$: $0.31 \lappeq \sin^2{\theta_{23}}\lappeq 0.72$ with central value around  $0.5$. The solar angle $\theta_{12}$ is large, $\sin^2{\theta_{12}}\sim 0.3$, but certainly not maximal (by about 5-6 $\sigma$ now \cite{lastSNO}). The third angle $\theta_{13}$, strongly limited mainly by the CHOOZ experiment, has at present a $3\sigma$ upper limit given by about $\sin^2{\theta_{13}}\lappeq 0.08$.

\begin{table*}[htb]
\begin{center}
\caption[]{Square mass differences and mixing angles\cite{one}}
\label{tab01}
\begin{tabular}{|c|c|c|c|}   
\hline  
& & & \\   
&{\tt lower limit} & {\tt best value} & {\tt upper limit}\\
&($3\sigma$)& & ($3\sigma$)\\
\hline
& & & \\
$(\Delta m^2_{sun})_{\rm LA}~(10^{-5}~{\rm eV}^2)$ & 5.4& 6.9&9.5\\
& & & \\
\hline
& & & \\
$\Delta m^2_{atm}~(10^{-3}~{\rm eV}^2)$ & 1.4& 2.6& 3.7\\
& & & \\
\hline
& & & \\
$\sin^2\theta_{12}$ & 0.23 & 0.30 &0.39\\
& & & \\
\hline
& & & \\
$\sin^2\theta_{23}$ & 0.31 & 0.52 &0.72\\
& & & \\
\hline
& & & \\
$\sin^2\theta_{13}$ & 0 & 0.006 &0.054\\
& & & \\
\hline
\end{tabular} 
\end{center}
\end{table*}

In spite of this experimental progress there are still many alternative routes in constructing models of neutrino masses. This variety is mostly due to the
considerable ambiguities that remain. First of all, it is essential to know whether
the LSND signal \cite{lsnd}, which has not been confirmed by KARMEN\cite{karmen} and is currently being double-checked by MiniBoone\cite{miniboone}, will be confirmed or will be excluded.  If LSND is right we probably
need at least four light neutrinos; if not we can do with only the three known ones, as we assume here in the following.  As
neutrino oscillations only determine mass squared differences a crucial missing input is the absolute scale of neutrino masses (within the existing limits from terrestrial experiments and cosmology \cite{Mainzetc},\cite{WMAPetc}).

The following experimental information on the absolute scale of neutrino masses is available. From the endpoint of tritium beta decay spectrum we have  an absolute upper limit of
2.2 eV (at 95\% C.L.) on the mass of "$\bar \nu_e$" \cite{Mainzetc}, which, combined with the observed oscillation
frequencies under the assumption of three CPT-invariant light neutrinos, represents also an upper bound on the masses of
all active neutrinos. Complementary information on the sum of neutrino masses is also provided by the galaxy power
spectrum combined with measurements of the cosmic  microwave background anisotropies. According to the recent analysis of
the WMAP collaboration \cite{WMAPetc}, 
$\sum_i \vert m_i\vert < 0.69$ eV (at 95\% C.L.).  More conservative analyses give $\sum_i \vert
m_i\vert < 1- 1.8$ eV, still much more restrictive than the laboratory bound. But, to some extent, the cosmological bound depends on a number of assumptions (or, in fashionable terms, priors). However, for 3 degenerate neutrinos of mass $m$, depending on our degree of confidence in the cosmological bound, we can be sure that $m \lappeq
0.23-0.3-0.6~eV$.
The discovery of $0\nu \beta \beta$ decay would be very important because it would directly establish lepton number violation and
the Majorana nature of $\nu$'s, and provide direct information on the absolute
scale of neutrino masses. 
The present limit from $0\nu \beta \beta$ is $\vert m_{ee}\vert< 0.2$ eV or to be
more conservative
$\vert m_{ee}\vert < (0.3\div 0.5)$ eV, where $ m_{ee} =  \sum{U_{ei}^2 m_i}$ in terms of the mixing matrix and the mass eigenvalues (see eq.(\ref{3nu1gen}). 

Given that neutrino masses are certainly extremely
small, it is really difficult from the theory point of view to avoid the conclusion that L conservation is probably  violated.
In fact, in terms of lepton number violation the smallness of neutrino masses can be naturally explained as inversely proportional
to the very large scale where L is violated, of order $M_{GUT}$ or even $M_{Pl}$.
Given for granted that neutrinos are Majorana particles, their masses can arise either from the see-saw mechanism or from generic dimension-five non renormalizable operators \cite{two} of the form: 
\beq 
O_5=\frac{(H l)^T_i \lambda_{ij} (H l)_j}{M}+~h.c.~~~,
\label{O5}
\eeq  
with $H$ being the ordinary Higgs doublet, $l_i$ the SU(2) lepton doublets, $\lambda$ a matrix in  flavour space,
$M$ a large scale of mass and a charge conjugation matrix $C$
between the lepton fields is understood. 
Neutrino masses generated by $O_5$ are of the order
$m_{\nu}\approx v^2/M$ for $\lambda_{ij}\approx {\rm O}(1)$, where $v\sim {\rm O}(100~\GeV)$ is the vacuum
expectation value of the ordinary Higgs. Apriori comparable masses are obtained in the see-saw mechanism: the resulting neutrino mass matrix reads:
\beq  m_{\nu}=m_D^T M^{-1}m_D~~~.
\eeq 
that is the light neutrino masses are quadratic in the Dirac
masses and inversely proportional to the large Majorana mass.  For
$m_{\nu}\approx \sqrt{\Delta m^2_{atm}}\approx 0.05$ eV and 
$m_{\nu}\approx m_D^2/M$ with $m_D\approx v
\approx 200~GeV$ we find $M\approx 10^{15}~GeV$ which indeed is an impressive indication for
$M_{GUT}$. Thus neutrino masses are a probe into the physics at $M_{GUT}$. 

\section{The $\nu$-Mixing Matrix}

If we take maximal $s_{23}$  ($s_{ij}= \sin  \theta_{ij} $) and keep only linear terms in $u=  s_{13}e^{i\varphi}$, from experiment we find the following
structure of the
$U_{fi}$ ($f=e$,$\mu$,$\tau$, $i=1,2,3$) mixing matrix, apart from sign convention redefinitions: 
\bea  
&&U_{fi}= \nonumber\\
&&\left(\matrix{ c_{12}&s_{12}&u \cr  -(s_{12}+c_{12}u^*)/\sqrt{2}&(c_{12}-s_{12}u^*)/\sqrt{2}&1/\sqrt{2}\cr
(s_{12}-c_{12}u^*)/\sqrt{2}&-(c_{12}+s_{12}u^*)/\sqrt{2}&1/\sqrt{2}     } 
\right),\label{ufi1}\\
\eea
Even for three neutrinos the pattern of the neutrino mass spectrum is still undetermined: it can be approximately degenerate, or of the inverse hierarchy type or normally hierarchical. Given the observed frequencies and  the notation $\Delta m^2_{sun}\equiv \vert\Delta m^2_{12}\vert$,
$\Delta m^2_{atm}\equiv \vert\Delta m^2_{23}\vert$ with
$\Delta m^2_{12}=\vert m_2\vert^2-\vert m_1\vert^2 > 0$ and $\Delta m^2_{23}= m_3^2-\vert m_2\vert ^2$, the three possible patterns of mass eigenvalues are:
\bea 
&&{\tt{Degenerate}}:  |m_1|\sim |m_2| \sim |m_3|\gg |m_i-m_j|\nonumber\\ 
&&{\tt{Inverted~hierarchy}}:  |m_1|\sim
|m_2| \gg |m_3| \nonumber\\ 
&&{\tt{Normal~hierarchy} }:  |m_3| \gg |m_{2,1}\label{abc}
\eea  
Models based on all these patterns have been proposed and studied and all are in fact viable at present.

The detection of neutrino-less double beta decay would offer a way to possibly disintangle the 3 cases.  The quantity which is bound by experiments
is the 11 entry of the
$\nu$ mass matrix, which in general, from $m_{\nu}=U^* m_{diag} U^\dagger$, is given by :
\bea 
\vert m_{ee}\vert~=\vert(1-s^2_{13})~(m_1 c^2_{12}~+~m_2 s^2_{12})+m_3 e^{2 i\phi} s^2_{13}\vert,
\label{3nu1gen}
\eea
Starting from this general formula it is simple to
derive the following bounds for degenerate, inverse hierarchy or normal hierarchy mass patterns,.
\begin{itemize}
\item[a)]  Degenerate case. If $|m|$ is the common mass and we take $s_{13}=0$, which is a safe
approximation in this case, because $|m_3|$ cannot compensate for the smallness of $s_{13}$, we have
$m_{ee}\sim |m|(c_{12}^2\pm s_{12}^2)$.  Here the phase ambiguity has been reduced to a sign ambiguity which is sufficient
for deriving bounds.  So, depending on the sign we have
$m_{ee}=|m|$ or
$m_{ee}=|m|cos2\theta_{12}$. We conclude that in  this case $m_{ee}$ could be as large as the present experimental limit
but should be at least of
order $O(\sqrt{\Delta m^2_{atm}})~\sim~O(10^{-2}~ {\rm eV})$ unless the solar angle is practically maximal, in which case
the minus sign option can be arbitrarily small. But the experimental 2-$\sigma$ range of the solar angle does not
favour a cancellation by more than a factor of about 3.
\item[b)]  Inverse hierarchy case. In this case the same approximate formula $m_{ee}=|m|(c_{12}^2\pm s_{12}^2)$ holds 
because $m_3$ is small and $s_{13}$ can be neglected. The difference is that here we know that $|m|\approx 
\sqrt{\Delta m^2_{atm}}$ so that $\vert m_{ee}\vert<\sqrt{\Delta m^2_{atm}}~\sim~0.05$ eV. At the same time,
since a full cancellation between the two contributions cannot take place, we expect 
$\vert m_{ee}\vert > 0.01$ eV.

\item[c)]  Normal hierarchy case. Here we cannot in general neglect the $m_3$ term. However in this case $\vert
m_{ee}\vert~\sim~
\vert\sqrt{\Delta m^2_{sun}}~ s_{12}^2~\pm~\sqrt{\Delta m^2_{atm}}~ s_{13}^2\vert$ and we have the bound 
$\vert m_{ee}\vert <$ a few $10^{-3}$ eV.
\end{itemize}
Recently some evidence for $0\nu \beta \beta$ was claimed \cite{kla} corresponding to
$\vert m_{ee}\vert\sim (0.2\div 0.6)~ {\rm eV}$ ($(0.1\div 0.9)~ {\rm eV}$ in a more conservative estimate of the involved nuclear matrix elements). If confirmed this would rule out cases b) and c) and point to case a) or to models
with more than 3 neutrinos.

\section{ "Normal" versus "Special" Models}

After KamLAND, SNO and WMAP not too much hierarchy in neutrino masses is indicated by experiments: 
\bea
r = \Delta m_{sol}^2/\Delta m_{atm}^2 \sim 1/35.\label{r}
\eea
 Precisely at $3\sigma$: $0.018 \lappeq r \lappeq 0.053$. Thus, for a hierarchical spectrum, $m_2/m_3 \sim \sqrt{r} \sim 0.2$, which is comparable to the Cabibbo angle $\lambda_C \sim 0.22$ or $\sqrt{m_{\mu}/m_{\tau}} \sim 0.24$. This suggests that the same hierarchy parameters (raised to powers with o(1) exponents) are at work for quark, charged lepton and neutrino mass matrices. This in turn indicates that, in absence of some special dynamical reason, we do not expect a quantity like $\theta_{13}$ to be too small. Indeed it would be very important to know how small the mixing angle $\theta_{13}$  is and how close to maximal is $\theta_{23}$. Actually one can make a  distinction between "normal" and "special" models. For normal models  $\theta_{23}$ is not too close to maximal and $\theta_{13}$ is not too small, typically a small power of the self-suggesting order parameter $\sqrt{r}$, with $r=\Delta m_{sol}^2/\Delta m_{atm}^2 \sim 1/35$. Special models are those where some symmetry or dynamical feature assures in a natural way the near vanishing of $\theta_{13}$ and/or of $\theta_{23}- \pi/4$. Normal models are conceptually more economical and much simpler to construct. 
Typical categories of normal models are
\begin{itemize}
\item[a)] Anarchy: Models with approximately degenerate mass spectrum and no ordering principle, no approximate symmetry assumed in the neutrino mass sector \cite{anarchy} \cite{review}. The small value of r is accidental, due to random fluctuations of matrix elements in the Dirac and Majorana neutrino mass matrices. The see-saw formula being a product of 3 matrices generates a broad distribution of r values resulting from random input in each factor. All mixing angles are generically large: so we do not expect $\theta_{23}$ to be maximal nor $\theta_{13}$ to be too small.
\item[b)] Semianarchy: We have seen that anarchy is the absence of structure in the neutrino sector. Here we consider an attenuation of anarchy where
the absence of structure is limited to the 23 sector. The typical texture is in this case \cite{lopsu1}  \cite{review}: 
\beq m_\nu
\approx m
\left(
\begin{array}{ccc}
\delta& \epsilon&\epsilon \\
\epsilon& 1&1\\ \epsilon& 1& 1
\end{array}
\right)
\label{s-an}~~~,
\eeq
where $\delta$ and $\epsilon$ are small and by 1 we mean entries of $o(1)$ and also the 23 determinant is of $o(1)$. 
This texture can be realized for example with different U(1) charges for $(l_1, l_2,l_3)$ , eg $(a,0,0)$ appearing in the dim. 5 operator of eq.(\ref{O5}). Clearly, in general we would
expect two mass eigenvalues of order 1, in units of $m$, and one small, of order $\delta$ or $\epsilon^2$. 
This pattern does not fit the observed solar
and atmospheric observed frequencies. However, given that the ratio $r$ is not too small, we can assume that its small value is
generated accidentally, as for anarchy. We see that, if by chance the second eigenvalue $\eta\sim \sqrt{r}\sim \delta+\epsilon^2$, we can then obtain the correct value of $r$
with large but in general non maximal $\theta_{23}$ and $\theta_{12}$ and small $\theta_{13}\sim \epsilon$. The smallness of $\theta_{13}$ is the main advantage over anarchy, but the relation with $\sqrt{r}$ normally keeps $\theta_{13}$ not too small (eg $\delta\sim \epsilon^2$ in simple U(1) models).
\item[c)] In the limit of exact $L_e-L_{\mu}-L_{\tau}$ symmetry we have inverted hierarchy with $r=0$ and bi-maximal mixing (both  $\theta_{12}$ and  $\theta_{23}$ are maximal) \cite{invhier} \cite{review}. Simple forms of symmetry breaking cannot sufficiently displace $\theta_{12}$ from the maximal value because typically
$\tan^2{\theta_{12}} \sim 1+o(r)$. Viable normal models are obtained by arranging large contributions to $\theta_{23}$ and $\theta_{12}$  from the charged lepton mass diagonalization. But then, in order to obtain the measured value of $\theta_{12}$  the size of $\theta_{13}$ must be close to its present upper bound.
\item[d)] Normal hierarchy models with 23 determinant suppressed by see-saw \cite{review}: in the 23 sector one needs relatively large mass splittings to fit the small value of $r$ but nearly maximal mixing. This can be obtained if the 23 sub-determinant is suppressed by some dynamical trick. Typical examples are lopsided models \cite{lops1} \cite{lopsu1} \cite{lops2} with large off diagonal term in the Dirac matrices of charged leptons and/or neutrinos (in minimal SU(5) the d-quark and charged lepton mass matrices are one the transposed of the other, so that large left-handed mixings for charged leptons correspond to large unobservable right-handed mixings for d-quarks). Another typical example is the dominance in the see-saw formula of a small eigenvalue in $M_{RR}$, the right-handed Majorana neutrino mass matrix \cite{king}. When the 23 determinant suppression is implemented in a 33 context, normally $\theta_{13}$ is not protected from contributions that vanish with the 23 determinant, hence with $r$.
\end{itemize}

The fact that some neutrino mixing angles are large and even
nearly maximal, while surprising at the start, was eventually found to be well compatible with a unified picture of quark
and lepton masses within GUTs. The symmetry group at
$M_{GUT}$ could be either (SUSY) SU(5) or SO(10)  or a larger group. For example, normal models based on anarchy, semianarchy, inverted hierarchy or normal hierarchy can all be naturally
implemented  by simple assignments of U(1)$_{\rm F}$ horizontal charges in a semiquantitative unified
description of all quark and lepton masses in SUSY SU(5)$\times$ U(1)$_{\rm F}$. Actually, in this context, if one adopts
a statistical criterium, hierarchical models appear to be preferred over anarchy and among them normal hierarchy appears the most likely.

In conclusion we expect that experiment will eventually find that  $\theta_{13}$ is not too small and that  $\theta_{23}$ is sizably not maximal. But if, on the contrary, either $\theta_{13}$ very small or  $\theta_{23}$ very close to maximal will emerge from experiment or both, then theory will need to cope with this fact. One can imagine other types of special  models, for example one where, starting from the lagrangian basis where the symmetries of the model are specified, all neutrino mixings arise from the diagonalisation of the charged lepton mass matrix. In ref. \cite{chlep} we argue that, in presence of two large mixing angles, this dominance of charged lepton mass diagonalization does not "normally" happen, although we can devise special tricks to enforce this possibility. In particular we have constructed an example which is natural in the technical sense and moreover has a very small $\theta_{13}$, so that it is a special model also in this respect. It is  interesting to conceive and explore dynamical structures that could lead to special models in a natural way. 
Normal models have been extensively discussed in the literature \cite{review}, so we concentrate here on examples of special models.

\section{Some Special Models}

We want to discuss here some particularly special models where both $\theta_{13}$ and $\theta_{23}- \pi/4$ exactly vanish
\footnote{More precisely,
they vanish in a suitable limit, with correction terms that can be made negligibly small.}. Then the neutrino mixing matrix $U_{fi}$ ($f=e$,$\mu$,$\tau$, $i=1,2,3$), in the basis of diagonal charged leptons, is given by, apart from sign convention redefinitions: 
\begin{equation}  
U_{fi}= 
\left(\matrix{ c_{12}&s_{12}&0 \cr  -s_{12}/\sqrt{2}&c_{12}/\sqrt{2}&-1/\sqrt{2}\cr
-s_{12}/\sqrt{2}&c_{12}/\sqrt{2}&1/\sqrt{2}     } \right) ~~~~~,
\label{ufi1}
\end{equation} 
where $c_{12}$ and $s_{12}$ stand for $\cos{\theta_{12}}$ and $\sin{\theta_{12}}$, respectively. It is much simpler to write natural models of this type with $s_{12}$ small and thus many such attempts are present in the early literature. More recently, given the experimental value of $\theta_{12}$, the more complicated case of $s_{12}$ large was also attacked, using non abelian symmetries, either continuous or discrete 
\cite{max1,max2,max3,max4,hps,max5,malast}. In the flavour basis the general form of the neutrino mass matrix for $\theta_{13}=0$ (no CP violation!) and $\theta_{23}$ maximal is given by:
\begin{equation}  
m_{\nu}= 
\left(\matrix{ x&y&y\cr y&z&w\cr
y&w&z   } \right) ~~~~~,
\label{grimus}
\end{equation} 
In eq. (\ref{grimus}) 4 real parameters appear corresponding to 3 eigenvalues plus $\theta_{12}$.
In many examples the invoked symmetries are particularly ad hoc and/or no sufficient attention is devoted to corrections from higher dimensional operators that can spoil the pattern arranged at tree level and to the highly non trivial vacuum alignment problems that arise if naturalness is required also at the level of vacuum expectation values (VEVs). 

An interesting special case of eq. (\ref{ufi1}) is obtained 
for  $s_{12}=1/\sqrt{3}$, i.e. the so-called tri-bimaximal or Harrison-Perkins-Scott mixing pattern  (HPS) 
\cite{hps}, with the entries in the second column all equal to $1/\sqrt{3}$ in absolute value:
\begin{equation}
U_{HPS}= \left(\matrix{
\dd\sqrt{\frac{2}{3}}&\dd\frac{1}{\sqrt 3}&0\cr
-\dd\frac{1}{\sqrt 6}&\dd\frac{1}{\sqrt 3}&-\dd\frac{1}{\sqrt 2}\cr
-\dd\frac{1}{\sqrt 6}&\dd\frac{1}{\sqrt 3}&\dd\frac{1}{\sqrt 2}}\right)~~~~~. 
\label{2}
\end{equation}
This matrix is a good approximation to present data
\footnote{In the HPS scheme $\tan^2{\theta_{12}}= 0.5$, to be compared with the latest experimental
determination \cite{lastSNO}: $\tan^2{\theta_{12}}= 0.45^{+0.09}_{-0.08}$.}. This is a most special model where not only $\theta_{13}$ and $\theta_{23}- \pi/4$ vanish but also $\theta_{12}$ assumes a particular value. Clearly, in a natural realization of this model, a very constraining and predictive dynamics must be underlying.  

Interesting ideas on how to obtain the HPS mixing matrix have been discussed in refs \cite{hps}. 
The most attractive models 
are based on the discrete symmetry $A_4$, which appears as particularly suitable for the purpose, and were presented 
in ref. \cite{max3,malast}. Here we discuss some general features of HPS models and present our version of an $A_4$ model \cite{ourlast}. There are a number of 
substantial improvements in our version with respect to Ma in ref. \cite{malast}. First, the HRS matrix is 
exactly obtained in a first approximation when higher dimensional operators are neglected, without imposing ad 
hoc relations among parameters (in ref. \cite{malast}. the equality of $b$ and $c$ is not guaranteed by the symmetry).
The observed hierarchy of charged lepton masses is obtained by assuming a larger flavour symmetry. 
The crucial issue of the required VEV alignment in the scalar sector is considered with special attention and a natural 
solution of this problem was presented. We also keep the flavour scalar fields distinct from the normal Higgs bosons 
(a proliferation of Higgs doublets is disfavoured by coupling unification) and singlets under the Standard Model 
gauge group. Last not least, we study the corrections from higher dimensionality operators allowed by the 
symmetries of the model and  discuss the conditions on the cut-off scales and the VEVs in order for these 
corrections to be completely under control. 
%
%
\section{Basic Structure of the Model}
The HPS mixing matrix implies that in a basis where charged lepton masses are 
diagonal the effective neutrino mass matrix is given by $m_{\nu} = U_{HPS}  \rm{diag}(m_1,m_2,\\m_3)U_{HPS}^T$:
\begin{equation}
m_{\nu}=  \left[\frac{m_3}{2}\left(\matrix{
0&0&0\cr
0&1&-1\cr
0&-1&1}\right)+\frac{m_2}{3}\left(\matrix{
1&1&1\cr
1&1&1\cr
1&1&1}\right)+\frac{m_1}{6}\left(\matrix{
4&-2&-2\cr
-2&1&1\cr
-2&1&1}\right)\right]~~~~~. 
\label{1}
\end{equation}
The eigenvalues of $m_{\nu}$ are $m_1$, $m_2$, $m_3$ with eigenvectors $(-2,1,1)/\sqrt{6}$, $(1,1,1)/\\\sqrt{3}$ and $(0,1,-1)/\sqrt{2}$, respectively. In general, apart from phases, there are six parameters in a real symmetric matrix like $m_{\nu}$: here only three are left after the values of the three mixing angles have been fixed \`a la HPS. For a hierarchical spectrum $m_3>>m_2>>m_1$, $m_3^2 \sim \Delta m^2_{atm}$, $m_2^2/m_3^2 \sim \Delta m^2_{sol}/\Delta m^2_{atm}$ and $m_1$ could be negligible. But also degenerate masses and inverse hierarchy can be reproduced: for example, by taking $m_3= - m_2=m_1$  we have a degenerate model, while for $m_1= - m_2$ and $m_3=0$ an inverse hierarchy case (stability under renormalization group running strongly prefers opposite signs for the first and the second eigenvalue which are related to solar oscillations and have the smallest mass squared splitting). From the general expression of the eigenvectors one immediately sees that this mass matrix, independent of the values of $m_i$, leads to the HPS mixing matrix. It is a curiosity that the eigenvectors are the same as in the case of the Fritzsch-Xing (FX) matrix 
\cite{FX} but with the roles of the first and the third ones interchanged (so that for HPS $\theta_{23}$ is maximal while $\sin^2{2\theta_{12}}=8/9$, while for FX the two mixing angles keep the same values but are interchanged). 

In ref. \cite{ourlast} (see also  \cite{nogo}) we show that if we want to reproduce $\theta_{23}=\pi/4$ in some limit
of our theory, necessarily this limit cannot correspond
to an exact symmetry in flavour space (we explicitly exclude symmetries that are broken by o(1) terms, for example such that the difference between the $\mu$ and the $\tau$ masses is a breaking effect or is introduced by hand while the symmetry would prescribe them of the same order). Then a maximal atmospheric
mixing angle can only originate from breaking effects
as a solution of a vacuum alignment problem.

Our model is based on the discrete group $A_4$ following refs \cite{max3,malast}, where its structure and representations are described in detail. Here we simply recall that $A_4$ is the discrete symmetry group of the rotations that leave a tethraedron invariant, or the group of the even permutations of 4 objects. It has 12 elements and 4 inequivalent irreducible representations denoted 1, $1'$, $1''$ and 3  in terms of their respective dimensions. Introducing $\omega$, the cubic root of unity, $\omega=\exp{i\frac{2\pi}{3}}$, so that $1+\omega+\omega^2=0$, the three one-dimensional representations are obtained by dividing the 12 elements of $A_4$ in three classes, which are determined by the multiplication rule, and assigning to (class 1, class 2, class 3) a factor $(1,1,1)$ for 1, or $(1,\omega,\omega^2)$ for $1'$ or $(1,\omega^2,\omega)$ for $1''$. The product of two 3 gives $3 \times 3 = 1 + 1' + 1'' + 3 + 3$. Also $1' \times 1' = 1''$, $1' \times 1'' = 1$, $1'' \times 1'' = 1'$ etc.
For $3\sim (a_1,a_2,a_3)$, $3'\sim (b_1,b_2,b_3)$ the irreducible representations obtained from their product are:
\begin{equation}
1=a_1b_1+a_2b_2+a_3b_3
\end{equation}
\begin{equation}
1'=a_1b_1+\omega a_2b_2+\omega^2 a_3b_3
\end{equation}
\begin{equation}
1''=a_1b_1+\omega^2 a_2b_2+\omega a_3b_3
\end{equation}
\begin{equation}
3\sim (a_2b_3, a_3b_1, a_1b_2)
\end{equation}
\begin{equation}
3\sim (a_3b_2, a_1b_3, a_2b_1)
\end{equation}
Following ref. \cite{malast} we assigns leptons to the four inequivalent
representations of $A_4$: left-handed lepton doublets $l$ transform
as a triplet $3$, while the right-handed charged leptons $e^c$,
$\mu^c$ and $\tau^c$ transform as $1$, $1'$ and $1''$, respectively. 
The flavour symmetry is broken by two real triplets
$\varphi$ and $\varphi'$ and by a real singlet $\xi$. 
At variance with the choice made by \cite{malast}, these fields 
are gauge singlets.
Hence we only need two Higgs doublets $h_{u,d}$ (not three generations of
them as in ref. \cite{malast}), which we take invariant under $A_4$. 
We assume that some mechanism produces and maintains the hierarchy
$\langle h_{u,d}\rangle=v_{u,d}\ll \Lambda$ where $\Lambda$ is the 
cut-off scale of the theory
\footnote{This is the well known hierarchy 
problem that can be solved, for instance, by realizing a supersymmetric 
version of this model.}.
The Yukawa interactions in the lepton sector read:
\be
{\cal L}_Y=y_e e^c (\varphi l)+y_\mu \mu^c (\varphi l)''+
y_\tau \tau^c (\varphi l)'+ x_a\xi (ll)+x_d (\varphi' ll)+h.c.+...
\label{wl}
\ee
In our notation, $(3 3)$ transforms as $1$, 
$(3 3)'$ transforms as $1'$ and $(3 3)''$ transforms as $1''$.
Also, to keep our notation compact, we use a two-component notation
for the fermion fields and we set to 1 the Higgs fields
$h_{u,d}$ and the cut-off scale $\Lambda$. For instance 
$y_e e^c (\varphi l)$ stands for $y_e e^c (\varphi l) h_d/\Lambda$,
$x_a\xi (ll)$ stands for $x_a\xi (l h_u l h_u)/\Lambda^2$ and so on.
The Lagrangian  ${\cal L}_Y$ contains the lowest order operators
in an expansion in powers of $1/\Lambda$. Dots stand for higher
dimensional operators. 
Some terms allowed by the flavour symmetry, such as the terms 
obtained by the exchange $\varphi'\leftrightarrow \varphi$, 
or the term $(ll)$ are missing in ${\cal L}_Y$. 
Their absence is crucial and is guaranteed by an additional discrete
$Z_4$ symmetry under which $f^c$ transform
into $-i f^c$ $(f=e,\mu,\tau)$, $l$ into $i l$, $\varphi$ is invariant and $\varphi'$
changes sign. This symmetry also explains why $\varphi$ and $\varphi'$ cannot be interchanged.

We need a mechanism such that the fields $\varphi'$,
$\varphi$ and $\xi$ develop a VEV along the directions:
\bea
\langle \varphi' \rangle&=&(v',0,0)\nn\\ 
\langle \varphi \rangle&=&(v,v,v)\nn\\
\langle \xi \rangle&=&u~~~. 
\label{align}
\eea 
Then at the leading order of the $1/\Lambda$ expansion,
the mass matrices $m_l$ and $m_\nu$ for charged leptons and 
neutrinos are given by:
\be
m_l=v_d\frac{v}{\Lambda}\left(
\begin{array}{ccc}
y_e& y_e& y_e\\
y_\mu& y_\mu \omega& y_\mu \omega^2\\
y_\tau& y_\tau \omega^2& y_\tau \omega
\end{array}
\right)~~~,
\label{mch}
\ee
\be
m_\nu=\frac{v_u^2}{\Lambda}\left(
\begin{array}{ccc}
a& 0& 0\\
0& a& d\\
0& d& a
\end{array}
\right)~~~,
\label{mnu}
\ee
where
\be
a\equiv x_a\frac{u}{\Lambda}~~~,~~~~~~~d\equiv x_d\frac{v'}{\Lambda}~~~.
\label{ad}
\ee
Charged leptons are diagonalized by
\be
l\to \frac{1}{\sqrt{3}}\left(
\begin{array}{ccc}
1& 1& 1\\
1& \omega^2& \omega\\
1& \omega& \omega^2
\end{array}
\right)l~~~,
\label{change}
\ee
and charged fermion masses are given by:
\be
m_e=\sqrt{3} y_e v_d \frac{v}{\Lambda}~~~,~~~~~~~
m_\mu=\sqrt{3} y_\mu v_d \frac{v}{\Lambda}~~~,~~~~~~~
m_\tau=\sqrt{3} y_\tau v_d \frac{v}{\Lambda}~~~.
\label{chmasses}
\ee
We can easily obtain a natural hierarchy among $m_e$, $m_\mu$ and
$m_\tau$ by introducing an additional U(1)$_F$ flavour symmetry under
which only the right-handed lepton sector is charged.
We assign F-charges $0$, $2$ and $3\div 4$ to $\tau^c$, $\mu^c$ and
$\e^c$, respectively. By assuming that a flavon $\theta$, carrying
a negative unit of F, acquires a VEV 
$\langle \theta \rangle/\Lambda\equiv\lambda<1$, the Yukawa couplings
become field dependent quantities $y_{e,\mu,\tau}=y_{e,\mu,\tau}(\theta)$
and we have
\be
y_\tau\approx O(1)~~~,~~~~~~~y_\mu\approx O(\lambda^2)~~~,
~~~~~~~y_e\approx O(\lambda^{3\div 4})~~~.
\ee
In the flavour basis the neutrino mass matrix reads 
\footnote{Notice that a unitary change of basis like the one in eq. (\ref{change})
will in general change the relative phases of the eigenvalues of $m_\nu$.}:
\be
m_\nu=\frac{v_u^2}{\Lambda}\left(
\begin{array}{ccc}
a+2 d/3& -d/3& -d/3\\
-d/3& 2d/3& a-d/3\\
-d/3& a-d/3& 2 d/3
\end{array}
\right)~~~,
\label{mnu0}
\ee
and is diagonalized by the transformation:
\be
U^T m_\nu U =\frac{v_u^2}{\Lambda}{\tt diag}(a+d,a,-a+d)~~~,
\ee
with
\be
U=\left(
\begin{array}{ccc}
\sqrt{2/3}& 1/\sqrt{3}& 0\\
-1/\sqrt{6}& 1/\sqrt{3}& -1/\sqrt{2}\\
-1/\sqrt{6}& 1/\sqrt{3}& +1/\sqrt{2}
\end{array}
\right)~~~.
\ee
The leading order predictions are $\tan^2\theta_{23}=1$, 
$\tan^2\theta_{12}=0.5$ and $\theta_{13}=0$. The neutrino masses
are $m_1=a+d$, $m_2=a$ and $m_3=-a+d$, in units of $v_u^2/\Lambda$.
We can express $|a|$, $|d|$ in terms of 
$r\equiv \Delta m^2_{sol}/\Delta m^2_{atm}
\equiv (|m_2|^2-|m_1|^2)/|m_3|^2-|m_1|^2)$,
$\Delta m^2_{atm}\equiv|m_3|^2-|m_1|^2$ 
and $\cos\Delta$, $\Delta$ being the phase difference between
the complex numbers $a$ and $d$:
\bea
\sqrt{2}|a|\frac{v_u^2}{\Lambda}&=&
\frac{-\sqrt{\Delta m^2_{atm}}}{2 \cos\Delta\sqrt{1-2r}}\nn\\
\sqrt{2}|d|\frac{v_u^2}{\Lambda}&=&
\sqrt{1-2r}\sqrt{\Delta m^2_{atm}}~~~.
\label{tuning}
\eea
To satisfy these relations a moderate tuning is needed in our model.
Due to the absence of $(ll)$ in eq. (\ref{wl}) which we will motivate in the next section, $a$ and $d$ are of the same order in $1/\Lambda$, 
see eq. (\ref{ad}). Therefore we expect that $|a|$ and $|d|$ 
are close to each other and, to satisfy eqs. (\ref{tuning}),
$\cos\Delta$ should be negative and of order one. We obtain:
\bea
|m_1|^2&=&\left[-r+\frac{1}{8\cos^2\Delta(1-2r)}\right]
\Delta m^2_{atm}\nn\\
|m_2|^2&=&\frac{1}{8\cos^2\Delta(1-2r)}
\Delta m^2_{atm}\nn\\
|m_3|^2&=&\left[1-r+\frac{1}{8\cos^2\Delta(1-2r)}\right]\Delta m^2_{atm}
\label{lospe}
\eea
If $\cos\Delta=-1$, we have a neutrino spectrum close to hierarchical:
\be
|m_3|\approx 0.053~~{\rm eV}~~~,~~~~~~~
|m_1|\approx |m_2|\approx 0.017~~{\rm eV}~~~.
\ee 
In this case the sum of neutrino masses is about $0.087$ eV.
If $\cos\Delta$ is accidentally small, the neutrino spectrum becomes
degenerate. The value of $|m_{ee}|$, the parameter characterizing the 
violation of total lepton number in neutrinoless double beta decay,
is given by:
\be
|m_{ee}|^2=\left[-\frac{1+4 r}{9}+\frac{1}{8\cos^2\Delta(1-2r)}\right]
\Delta m^2_{atm}~~~.
\ee
For $\cos\Delta=-1$ we get $|m_{ee}|\approx 0.005$ eV, at the upper edge of
the range allowed for normal hierarchy, but unfortunately too small
to be detected in a near future.
Independently from the value of the unknown phase $\Delta$
we get the relation:
\be
|m_3|^2=|m_{ee}|^2+\frac{10}{9}\Delta m^2_{atm}\left(1-\frac{r}{2}\right)~~~,
\ee
which is a prediction of our model.

It is also important to get some constraint on the mass scales involved
in our construction. From eqs. (\ref{tuning}) and (\ref{ad}), 
by assuming $x_d\approx 1$ $v_u\approx 250$ GeV, we have
\be
\Lambda\approx 1.8\times 10^{15}~\left(\frac{v'}{\Lambda}\right)~~{\rm GeV}~~~.
\ee
Since, to have a meaningful expansion, we expect $v'\le\Lambda$,
we have the upper bound
\be
\Lambda< 1.8\times 10^{15}~~{\rm GeV}~~~.
\label{comax}
\ee
Beyond this energy scale, new physics should come into play.
The smaller the ratio $v'/\Lambda$, 
the smaller becomes the cut-off scale.
For instance, when $v'/\Lambda=0.03$, $\Lambda$ should be close to $10^{14}$
GeV. A complementary information comes from the charged lepton sector,
eq. (\ref{chmasses}). A lower bound on $v/\Lambda$ can be derived
from the requirement that the Yukawa coupling $y_\tau$ remains in
a perturbative regime. By asking $y_\tau v_d< 250$ GeV, we get
\be
\frac{v}{\Lambda}>0.004~~~.
\ee
Finally, by assuming that all the VEVs fall in approximately the same 
range, which will be shown in section 5, we obtain the range
\be
0.004<\frac{v'}{\Lambda}\approx \frac{v}{\Lambda}\approx 
\frac{u}{\Lambda}<1~~~,
\label{vsuL}
\ee that will be useful to estimate the effects of higher-dimensional
operators.
Correspondingly the cut-off scale will range between
about $10^{13}$ and $1.8\times 10^{15}$ GeV.
\vspace{0.5cm}
%
%
\section{Vacuum alignment in a $A_4$ model in an extra dimension}
The problem of achieving the
vacuum alignment of eq. (\ref{align}) is not at all trivial. At the same time, to produce
the desired mass matrices in the neutrino and charged lepton
sectors, we
should prevent, at least at some level, the interchange
between the fields $\varphi$ and $\varphi'$ . There are several difficulties
to naturally accomplish these requirements. 
By minimizing the scalar potential of the theory with
respect to $\varphi$ and $\varphi'$ we get six equations
that we would like to satisfy in terms of the two unknown
$v$ and $v'$. Even though we expect that, due to the symmetry $A_4$, 
the six minimum conditions are not necessarily independent,
such an expectation turns out to be wrong in the specific case, unless some additional
relation is enforced on the parameters of the scalar potential.
These additional relations are in general not natural.
For instance, even by imposing them at the tree level,
they are expected to be violated at the one-loop order.
Therefore it turns out that without some special trick the minimum conditions
cannot be all satisfied by our vacuum configuration. We now discuss a solution to this problem.
\begin{center}
\begin{figure}
\vspace{6.0cm}
  \includegraphics{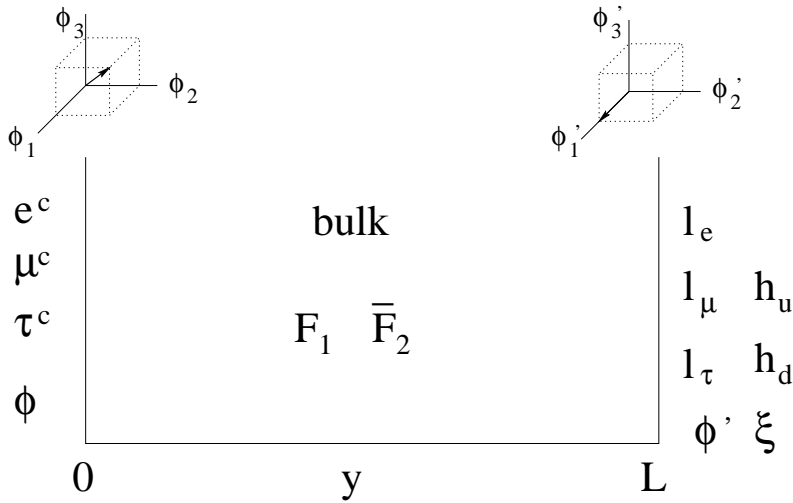}
   \caption{\it
   Fifth dimension and localization of scalar and fermion fields.
The symmetry breaking sector includes the $A_4$ triplets $\varphi$
and $\varphi'$, localized at the opposite ends of the interval.
Their VEVs are dynamically aligned along the directions shown
at the top of the figure.}
\end{figure}
\end{center}

One of the problems we should overcome in the search for
the correct alignment is that of keeping neutrino and charged
lepton sectors separate, including the respective symmetry 
breaking sectors. Here we show that such a separation can be achieved
by means of an extra spatial dimension. The space-time is assumed
to be five-dimensional, the product of the four-dimensional
Minkowski space-time times an interval going from $y=0$ to $y=L$. 
At $y=0$ and $y=L$ the space-time has two four-dimensional boundaries,
which we will call branes. Our idea is that matter SU(2) singlets
such as $e^c,\mu^c,\tau^c$ are localized at $y=0$, while SU(2) doublets,
such as $l$ are localized at $y=L$ (see Fig.1). Neutrino masses
arise from local operators at $y=L$. Charged lepton
masses are produced by non-local effects involving both branes.
The simplest possibility is to introduce a bulk fermion $F(x,y)$, depending on all space-time coordinates, that interacts with
$e^c,\mu^c,\tau^c$ at $y=0$ and with $l$ at $y=L$. The exchange of
such a fermion can provide the desired non-local coupling between
right-handed and left-handed ordinary fermions. We also impose the discrete $Z_4$
symmetry introduced in the previous section under which $(f^c,l,F,\varphi,
\varphi',\xi)$ transform into 
$(-i f^c,i l,i F,\varphi,-\varphi',-\xi)$,
Finally,
assuming that $\varphi$ and $(\varphi',\xi)$ are localized
respectively at $y=0$ and $y=L$, we obtain a natural separation
between the two sectors
and their respective scalar potentials are minimized by the 
desired field configurations, for natural values of the implied parameters.

%
Such a mechanism only works in the case of discrete symmetries,
since in the continuous case the large symmetry of the total
potential energy would make the relative
orientations of the two scalar sectors undetermined.

Last but not least, the hierarchy of the charged lepton masses
can be reproduced by the usual Froggatt-Nielsen mechanism
within the context of an abelian flavour symmetry, which
turns out to be fully compatible with the present scheme.

In ref. \cite{ourlast} we have extensively discussed how this lowest order picture is modified by
the introduction of higher dimensional operators.
The induced corrections are parametrically small, of second
order in the expansion parameter $VEV/\Lambda$, $\Lambda$ being the
cut-off of the theory, and they can be made
numerically negligible.

We believe that, from a purely technical point of view,  we have fulfilled our goal to realize a completely natural construction of the HPS mixing scheme. But to construct our model we had to introduce a number of special dynamical tricks (like a peculiar set of discrete symmetries in extra dimensions). Apparently this is  the price to pay for a ``special'' model where all mixing angles are fixed to particular values. Perhaps this exercise can be taken as a hint that it is more plausible to expect that, in the end, experiment will select a ``normal'' model with $\theta_{13}$ not too small and $\theta_{23}$ not too close to maximal.

\vskip 0.2 cm
\section*{Acknowledgment}
I thank Giorgio Bellettini, Giorgio Chiarelli and Mario Greco for their kind invitation to this perfectly organised Conference

\end{document}